\documentclass[10pt,a4paper]{article}

\usepackage{graphics}
\newcommand{\bea}{\begin{eqnarray}}
\newcommand{\eea}{\end{eqnarray}}
\newcommand{\be}{\begin{equation}}
\newcommand{\ee}{\end{equation}}

\def\be{\begin{eqnarray}}
\def\ee{\end{eqnarray}}
\def\bd{\begin{displaymath}}
\def\ed{\end{displaymath}}

\def\ga{\gamma}

\def\ADNDT{{At. Data Nucl. Data Tables }}

\def\NP{Nucl. Phys. }
\def\PR{Phys. Rev. }
\def\PRL{Phys. Rev. Lett. }

\def\PRep{Phys. Rep. }
\def\jpg{J. Phys. G: Nucl. Part. Phys. }
\def\EPJ{{Eur. Phys. J. }}
\def\IJMPE{{Int. J. Mod. Phys. E }}
\def\etal{{\em et al. }}
\def\g{{\gamma}}

\begin{document}
\title{Low energy proton reactions of astrophysical interest in A$\sim90-100$ region}

\author{C. Lahiri and G. Gangopadhyay\\
Department of Physics, University of Calcutta\\
92, Acharya Prafulla Chandra Road, Kolkata-700 009, India}

\date{}
\maketitle
\begin{abstract}
Semimicroscopic optical potentials for low energy proton reactions in mass 
$90-100$ region have been obtained by folding the density dependent M3Y 
interaction with relativistic mean field densities. 
Certain parameters in the potential have been deduced by comparing calculated results with the  data for elastic scattering. Low energy proton reactions 
in this mass region have been studied in the formalism with success.
Rates of important astrophysical reaction in the mass region have been calculated.
\end{abstract}

The $p$-process is a common term given to the astrophysical reactions, which are 
involved in synthesis of heavy elements but do not correspond to the $r$- or 
the $s$-processes.  
It includes reactions such as proton capture, charge 
exchange and photodisintegration. The $p$-process is known to be important for 
production of 
certain so called $p$-nuclei, which are beyond the ambit of the slow and fast 
neutron reactions. A $p$-network involves typically two 
thousand nuclei, and incorporates,  again typically, twenty thousand 
reactions and decays. More details may be found in standard text books [for 
example Illiadis\cite{book}] and reviews\cite{prev}.

One obvious problem in studying the $p$-process is that many of the involved 
nuclei have very short life times and  are not available in our terrestrial 
laboratories for experiment. Though radioactive ion beams 
have opened a new vista, we are still far away from having the reaction rates 
at astrophysical energies for all the main reactions involved in the p-process. 
Thus, theoretical calculations for such rates remain very important for 
the $p$-process. For example, Rapp \etal have identified a number of reactions,
which are very important in $p$-process\cite{astro}. In mass 90 -100 region, the
list includes the photodisintegration reactions emitting protons and leading 
to the products $^{91}$Nb, $^{95}$Tc, and $^{99}$Rh. The rates for these 
reactions could be obtained from their inverse, {\em i.e.} ($p,\gamma)$
reactions, had the above nuclei been easily available in laboratories.  

There have been numerous theoretical calculations of astrophysical rates\cite{rauser} 
employing various models. However, very often in literature,  these
theoretical rates are varied 
by factors ranging from ten to hundred to study their effects\cite{sch}.
In some earlier works, we calculated the cross sections of various low energy 
proton reactions some of which are involved in the rapid proton 
processes in mass 60-80 region\cite{prc1,epj,prc}. The 
semimicroscopic optical model was employed for calculation of cross sections
using densities from theoretical mean field calculations. It is our aim to 
fix the various parameters and prescriptions in our procedure by fitting 
available low energy cross sections for various 
reactions in a mass region and calculate the rates for various 
reactions involving protons, which are important in nucleosynthesis. Thus,
a more stringent restriction may be imposed on the variation of rates. 
Some consequences of the above approach in rapid proton process have already
been analyzed\cite{ijm,ijm1}. 

The code TALYS1.4\cite{talys} has been used to calculate cross sections and rates in the
 Hauser-Feshbach formalism. In our earlier works, it was concluded that the 
Hartree-Fock-Bogoliubov level densities, 
calculated in TALYS by Hilaire\cite{Hilaire}  and the E1 gamma strength functions, 
calculated in the same approach, fit the results in mass 60-80 region. In the present
calculation we employ these values to extend our calculations to mass 90 region.
 
The method followed in the present procedure has been detailed in our earlier
publications\cite{prc1,epj} and is not discussed here. The FSU Gold\cite{prl} Lagrangian 
density is employed to calculate the nuclear density. Since we need the density 
as a function of radius, the calculation is performed in the co-ordinate space.
We employ spherical approximation, as most the nuclei 
under study are near closed shells for both protons and neutrons and are not 
strongly deformed.
The effective interaction DDM3Y\cite{ddm3y}, derived 
from nuclear matter calculation, has been folded with the nuclear densities to 
obtain the semi-microscopic optical model potentials in the Local Density 
Approximation.

Since the nuclear density is an important factor in the present formalism, 
we have studied the charge radii values. The charge radius is the first order 
moment of the charge distribution. In Table \ref{radius}, we compare our results for the 
charge radii ($r_{ch}$) with measurements for those nuclei 
in this mass region, which have been involved in the reactions studied later 
in this work and for which experimental radius information are available. 
Charge densities have been obtained by folding point proton densities
with a Gaussian form factor to incorporate the effect of the finite size of the 
proton as in our previous work\cite{prc}. It is clear that the charge radii 
are reasonably well produced in our calculation.

\begin{figure}
\resizebox{!}{!}{\includegraphics{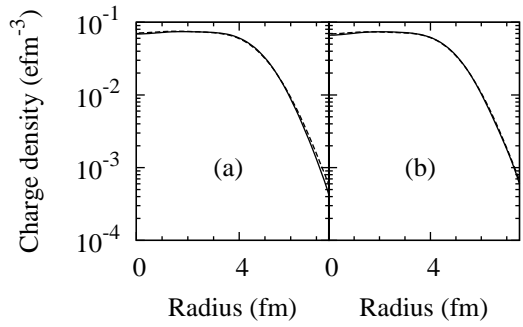}}
\caption{Comparison of charge density obtained from Fourier-Bessel analysis of 
experimental electron scattering data (solid line) and calculated in the present work (dashed line) for (a) $^{90}$Zr
and (b) $^{94}$Mo respectively.
\label{density}}
\end{figure}

We could not find direct experimental values for charge densities. Hence,
we have employed the Fourier-Bessel coefficients for densities extracted from 
electron scattering experiments in de Vries \etal\cite{chden} to get the charge 
densities and  plotted two examples in Fig. \ref{density}. 
One can see that the theoretical results reasonably agree with experiments. 
However, the absence of any information on error prevents us from reaching a 
firm conclusion.

As a first test of the optical model potential, we have looked at elastic 
proton scattering at low energies. Elastic scattering involves the same 
incoming and outgoing channel
for the optical model and may be taken to provide the simplest test to 
constrain various parameters involved in the calculation.
The proton energy relevant to a typical $p$-process temperature of 1$-$3 GK 
for nuclei in this mass region lies between 1$-$4 MeV.
However, scattering experiments are very difficult at such low energies,  as 
the cross sections are extremely small, and hence no
experimental data are available. We have compared the cross sections 
at the lowest energies available in literature with theoretical
results. 

\begin{figure}
\resizebox{8.5cm}{!}{\includegraphics{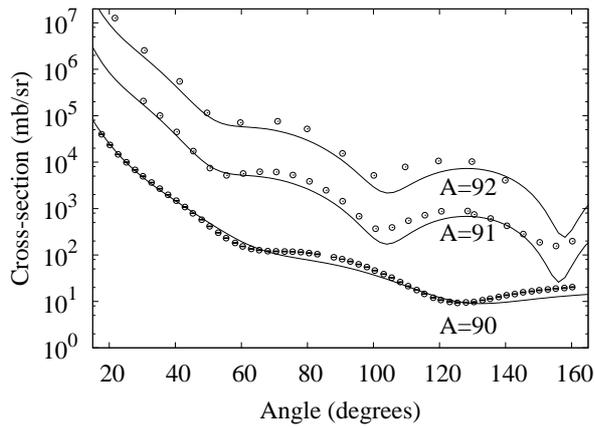}}
\caption{Experimental and calculated cross sections for elastic proton
scattering in Zr isotopes. For $A=90$, 91 and 92, the proton energies are
 9.7 MeV, 14.8 MeV,  and 14.25 MeV, respectively. The cross sections for
$^{91,92}$Zr have been multiplied by a factor of 100 and 1000, respectively.\label{Zrpel}}
\end{figure}

\begin{figure}
\resizebox{8.5cm}{!}{\includegraphics{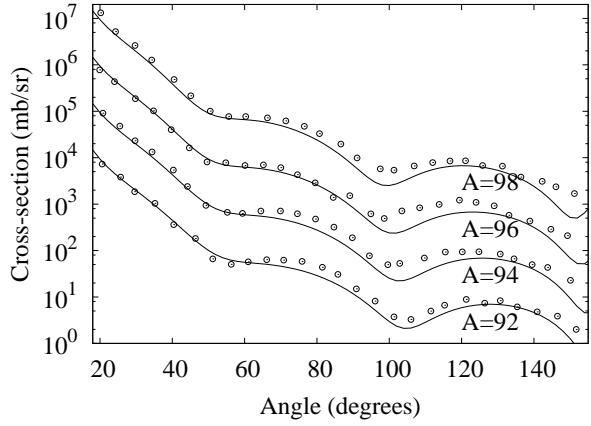}}
\caption{Experimental and calculated cross sections for elastic proton
scattering in Mo isotopes at 15 MeV proton energy. The cross sections for
$A=$ 92, 94, and 96 have been multiplied by factors of 10, 100, and 1000, respectively.\label{Mopel}}
\end{figure}

In Figs. \ref{Zrpel} and \ref{Mopel}, we present the results of some 
of our calculations in Zr and Mo isotopes, respectively, along with the corresponding experimental results. 
Experimental values are respectively from Refs. 
\cite{zr90elas,zr91elas,zr92elas} for $^{90,91,92}$Zr, and from Ref. 
\cite{moelas} for Mo isotopes.
To fit the experimental data, the 
folded DDM3Y potential has been multiplied by factors of 0.81 and 0.15 to 
obtain the real and imaginary parts of the optical potential, respectively. 
Throughout the rest of the work, we use these two factors to obtain the 
potential.  We emphasize that better fits for individual 
reactions are possible by varying different parameters.
But if the present calculation has to be extended to unknown mass region, this approach is
clearly inadequate. Therefore, we have refrained from fitting individual reactions. 
In our previous work\cite{epj,prc}, we used a different normalization which is in good agreement
with experimental values in a wide mass region(A $\approx$ 60-88). But beyond that region, same set of parameters 
are unable to fit the experimental data for p-nuclei\cite{astro} and therefore, we choose the above
set of parameters. Though, there are no sharp boundaries for a 
mass region, but for simplicity, we choose it in such a way that a single set of parameters can fit
the entire mass region. In present work, we have chosen the mass region A$\approx$89-100.    

It is clear from Figs. \ref{Zrpel} and \ref{Mopel} that the DDM3Y interaction
can describe the data well. In fact, we have found that as one goes to lower 
energies, the quality of agreement tends to improve. Thus at energies relevant 
to astrophysical interest, we can expect the present method 
to provide a good description. 

\begin{figure}
\resizebox{8.5cm}{!}{\includegraphics{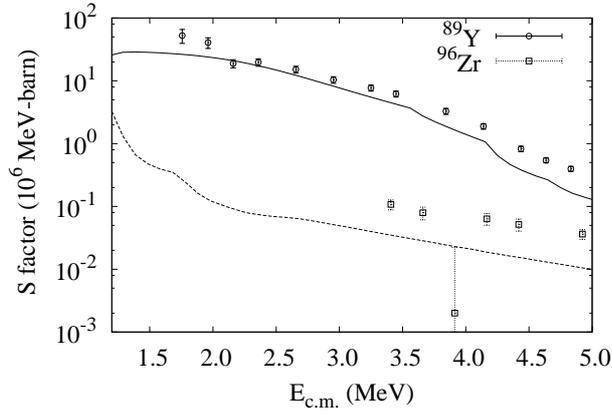}}
\caption{Experimental and calculated S-factors for ($p,\gamma$) reactions in
$^{89}$Y and $^{96}$Zr, respectively. The solid (dashed) line indicates 
calculated results
for $^{89}$Y($^{96}$Zr). \label{Y89}}
\end{figure}

\begin{figure}
\resizebox{8.5cm}{!}{\includegraphics{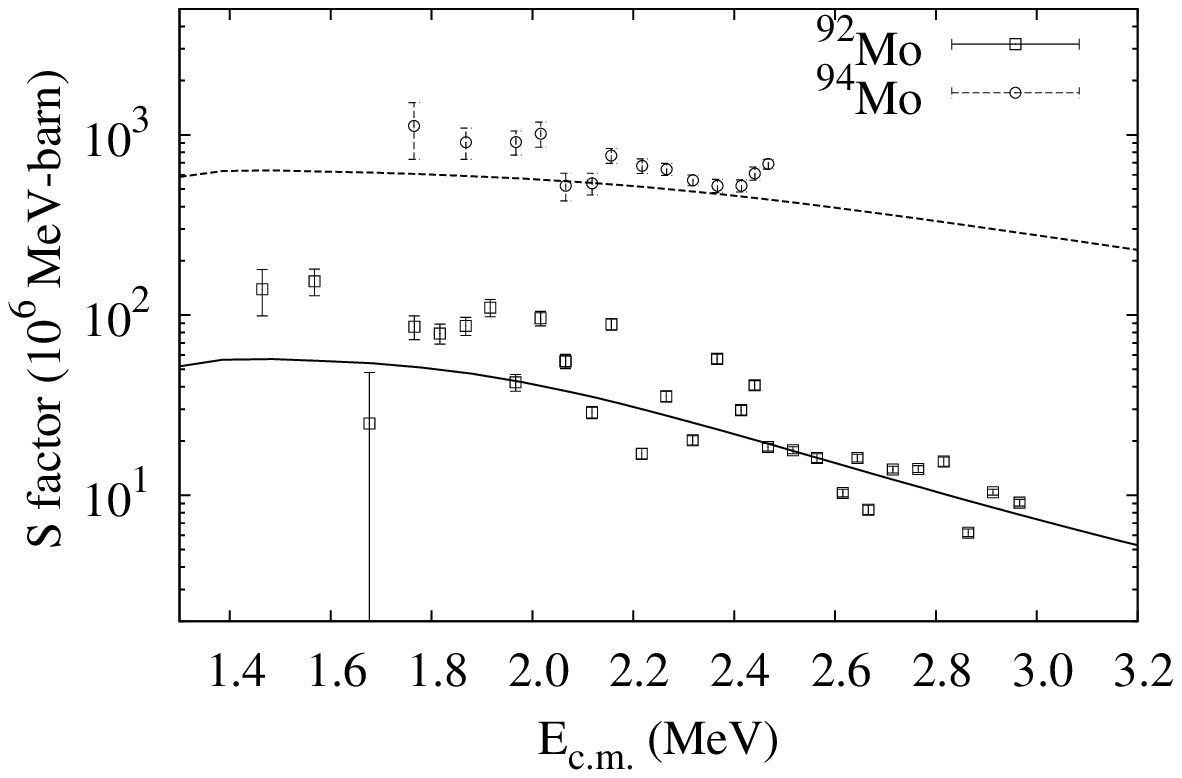}}
\caption{Experimental and calculated S-factors for $^{92,94}$Mo($p,\gamma$) 
reactions. Results for $^{94}$Mo have been multiplied by 10. The solid (dashed) line indicates calculated results
for $^{92}$Mo($^{94}$Mo).\label{Mo9294}}
\end{figure}

\begin{figure}
\resizebox{8.5cm}{!}{\includegraphics{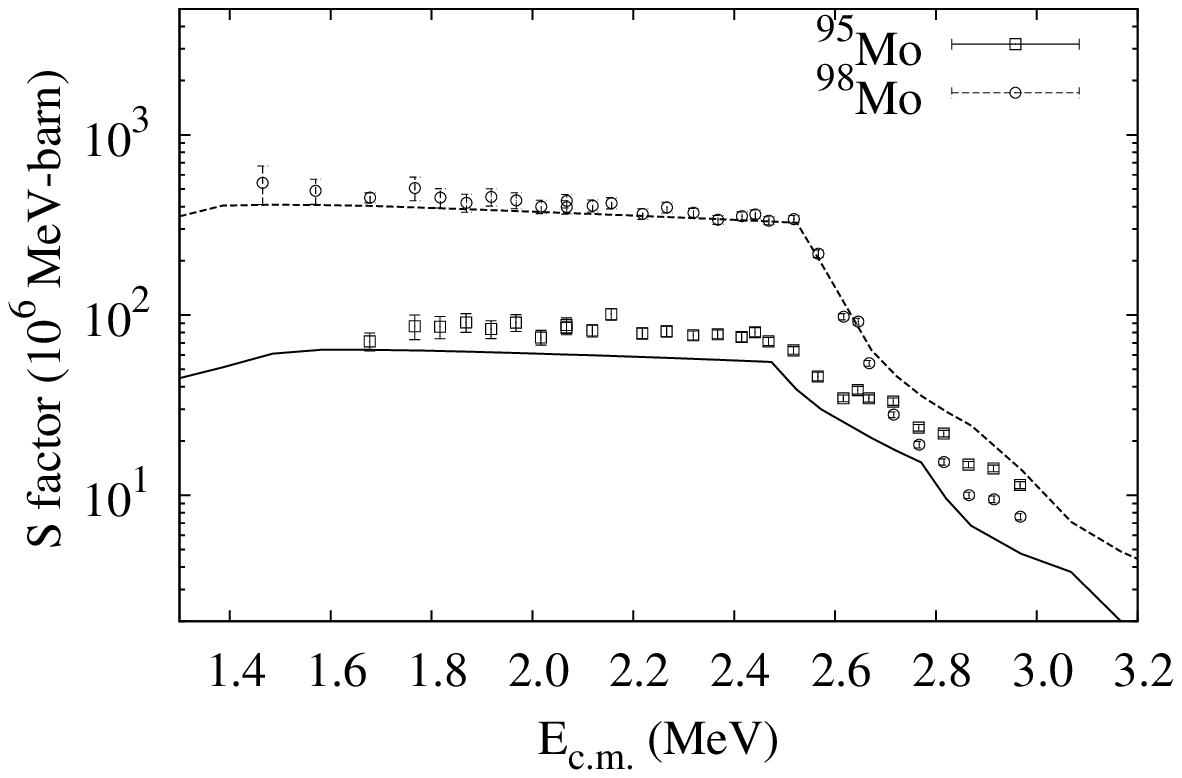}}
\caption{Experimental and calculated S-factors for $^{95,98}$Mo($p,\gamma$) 
reactions. Results for $^{98}$Mo have been multiplied by 10. The solid (dashed) line indicates calculated results
for $^{95}$Mo($^{98}$Mo). \label{Mo9598}}
\end{figure}

\begin{figure}
\resizebox{8.5cm}{!}{\includegraphics{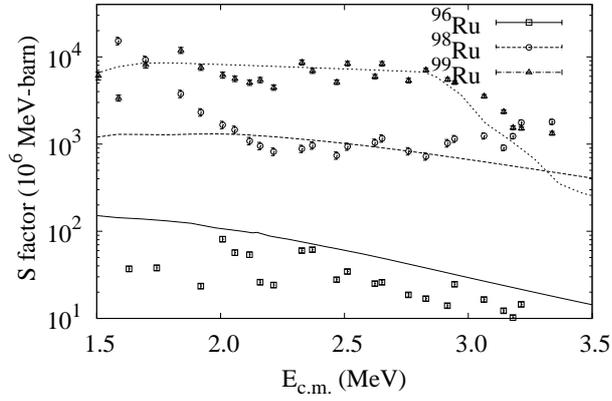}}
\caption{Experimental and calculated S-factors for $^{96,98,99}$Ru($p,\gamma$) 
reactions. Results for $^{98,99}$Ru have been multiplied by 10 and 100, 
respectively. The solid, dashed and dotted lines indicate results for 
$^{96,98,99}$Ru, respectively.\label{Ru9698}}
\end{figure}

The same formalism has been used to study the low energy $(p,\ga)$ 
reactions in a number of nuclei in this mass region. As the cross section 
varies very rapidly at low energies, it is more convenient to present 
the S-factor values. In Figs. \ref{Y89}-\ref{Ru9698}, calculated values are 
compared with experimental results. Next, we very briefly discuss our results.

For $^{89}$Y, the experimental values are from Tsagari \etal\cite{y89p}.
For $^{96}$Zr, the results are from Chloupek \etal\cite{zr96p}, though
there seems to be certain error in the values in that reference. The 
numerical values presented there are larger by a factor of 10$^3$
that the values presented in Fig. 9 of the reference. The latter values appear 
to be correct to the present authors and and are indicated in Fig. \ref{Y89}.
The data for Mo and Ru isotopes are from Ref \cite{mop,rup}, respectively.

For $^{96}$Zr target, there are very few experimental points within the energy 
range important for astrophysical reactions. 
In $^{89}$Y, the trend of the experimental values has been correctly reproduced.
None of the experimental values
differs by a more than a factor of two. The last two comments are generally valid for almost all the other reactions.
 One important exception is the
$^{98}$Ru$(p,\g)$ reaction, where the measured cross section 
systematically increases with decrease in energy below 2 MeV proton energy
compared to the calculated values and becomes larger by more than one order of 
magnitude around 1.6 MeV. It has not been possible to explain such a large 
increase, which is absent in all low energy $(p,\g)$ reactions in this mass 
region for which data are available. In fact, the calculation carried out in
Ref. \cite{rup}, where the experimental values have been published, is
also unable to explain such a sudden increase.
 
\begin{figure}
\resizebox{8.5cm}{!}{\includegraphics{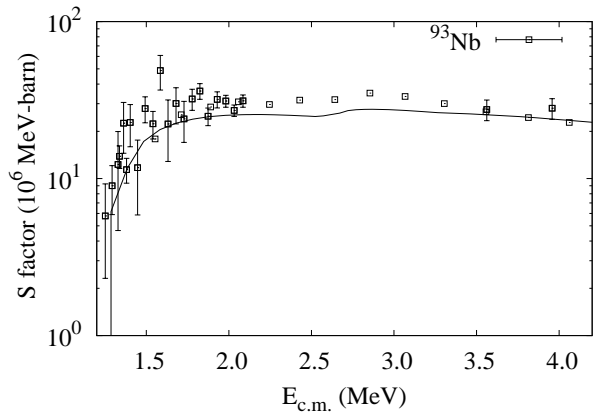}}
\caption{Experimental and calculated S-factors for the $^{93}$Nb($p,n$) 
reaction. \label{pn}}
\end{figure}

We next look for other low energy reactions 
involving proton projectile. The only  reaction for which we have been able to 
find substantial amount of  data in the domain of astrophysical 
energies is the $^{93}$Nb$(p,n)$ reaction\cite{nb93pn}. 
Our results have been presented in 
Fig. \ref{pn}. One can see that our calculation gives an excellent
description of the experimental trends. However, one should also note that the
data are rather old and have either very large errors or no quoted error value.

\begin{figure}
\resizebox{!}{!}{\includegraphics{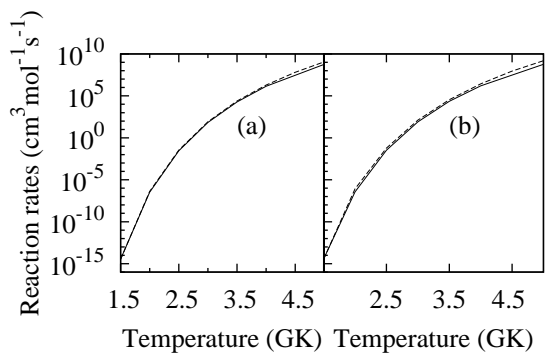}}
\caption{ Comparison of rates (cm$^3$ mol$^{-1}$ sec$^{-1}$) for $(\g,p)$ reactions from present 
calculation (solid line) and NON-SMOKER\cite{rauser} calculation (dotted line) for (a) $^{92}$Mo and
(b) $^{96}$Ru respectively.\label{ratefig}}
\end{figure}

For the sake of completeness, in Fig. \ref{ratefig} we compare the rates of $(\g,p)$ reaction from present calculation
 with rates from NON-SMOKER\cite{rauser}
calculation for $^{92}$Mo and $^{96}$Ru. One can see that the present calculation is very similar to the NON-SMOKER 
values. Therefore, it is expected that all the results can also be reproduced with commonly used NON-SMOKER rates.

From the above discussion, it is possible to conclude that the 
low energy reaction cross sections are, except in one case, 
reasonably reproduced in the above approach. The success in this calculation has
enabled us to calculate the astrophysical rates for the
reactions,  identified  as important by Rapp \etal\cite{astro}, in mass 90-100 
region. They are presented in Table \ref{rate}. 

To summarize, relativistic mean field calculation has been performed in nuclei between mass 
90 and 100 to obtain the density profiles. They, in turn, have been folded 
with the density dependent M3Y interaction to obtain the semimicroscopic 
optical potential.  Parameters in the potential have been fixed by comparing 
with low energy proton scattering. Available experimental information on low 
energy proton reactions has been compared with theory. Rates of important 
astrophysical reaction in the mass region have also been calculated.

This work has been carried out with financial assistance of the UGC sponsored
DRS Programme of the Department of Physics of the University of Calcutta. 
CL
acknowledges the grant of a fellowship awarded by the UGC.
GG acknowledges the facilities provided under the ICTP Associateship Programme by ECT*,
 Trento where a part of the work has been carried out. Discussion with Alexis 
Diaz-Torres is gratefully acknowledged.

\newpage

\begin{table}
\caption{Experimental charge radii values 
compared with calculated results for the nuclei involved in low energy proton reactions. The experimental values are from the compilation by Angeli\cite{radii}.
\label{radius}} 
\begin{center}
\begin{tabular}{lcc|lcc}\hline
&\multicolumn{2}{c|}{$r_{ch}$(fm)}& 
&\multicolumn{2}{c}{$r_{ch}$(fm)} \\
& Theo.
 & Exp.&&Theo.&Exp.\\\hline
$^{89}$Y & 4.274&4.242 &
$^{94}$Mo&4.366&4.352 \\ 
$^{90}$Zr&4.297&4.270  &  
$^{95}$Mo&4.376&4.362 \\ 
$^{92}$Zr& 4.317&4.306 &  
$^{96}$Mo&4.387&4.384 \\ 
$^{94}$Zr& 4.335&4.331 &  
$^{98}$Mo&4.407&4.409 \\ 
$^{96}$Zr&4.356&4.350 &  
$^{96}$Ru&4.409&4.393 \\ 
$^{92}$Mo&4.344&4.316&  
$^{98}$Ru&4.431&4.409 \\ 
\hline
\end{tabular}
\end{center}
\end{table}

\begin{table}
\caption{Rates in cm$^{3}$mole$^{-1}$sec$^{-1}$for selected $(\g,p)$ reactions of astrophysical importance.
\label{rate}}
\begin{tabular}{cccc}\hline
$T$ (GK) &\multicolumn{3}{c}{Target}\\
& $^{92}$Mo & $^{96}$Ru & $^{100}$Pd \\\hline
1.5&3.45$\times 10^{-15}$&3.67$\times 10^{-15}$&2.69$\times 10^{-14}$\\
2.0&3.34$\times 10^{-07}$&3.64$\times 10^{-07}$&1.52$\times 10^{-06}$\\
2.5&2.89$\times 10^{-02}$&3.31$\times 10^{-02}$&9.85$\times 10^{-02}$\\
3.0&6.77$\times 10^{+01}$&7.95$\times 10^{+01}$&1.85$\times 10^{+02}$\\
3.5&1.92$\times 10^{+04}$&2.23$\times 10^{+04}$&4.30$\times 10^{+04}$\\
4.0&1.40$\times 10^{+06}$&1.57$\times 10^{+06}$&2.60$\times 10^{+06}$\\
5.0&5.87$\times 10^{+08}$&5.70$\times 10^{+08}$&7.86$\times 10^{+08}$\\
\hline
\end{tabular}
\end{table}

\newpage



\end{document}